\documentclass[12pt]{article}
\usepackage{epsfig}
\usepackage{amsmath}
\usepackage{amsfonts}

\def\BE{\begin{equation}}
\def\EE{\end{equation}}
\def\BA{\begin{eqnarray}}
\def\EA{\end{eqnarray}}
\def\p{\partial}

\def\e{\epsilon}

\begin{document}
\begin{flushright}
DTP/99/33
\end{flushright}
\vspace{1.5cm} 
\begin{center}
{\large {\bf 
Calculating the infrared central charges for perturbed minimal models:
 improving the RG perturbation}}\\
\vspace{1cm}
{\large Lars Kj\ae rgaard and  Paul Mansfield} \\
\vspace{0.3cm}
{\small Department of Mathematical Sciences}\\
{\small University of Durham}\\ 
{\small South Road, Durham DH1 3LE, U.K.}\\
{\small \texttt{lars.kjaergaard@durham.ac.uk}} \\ 
{\small \texttt{p.r.w.mansfield@durham.ac.uk}} \\
\vspace{0.5cm}
\today
\end{center}
\vspace{0.5cm}

\begin{abstract}
\noindent 
We illustrate a method for improving Renormalisation Group
improved perturbation theory by calculating the infra-red
central charge of a perturbed conformal field theory.
The additional input is a dispersion relation that exploits
analyticity of the energy-momentum tensor correlator.
\end{abstract}

\section{Introduction}
The purpose of this letter is to describe a method for improving a
Renormalisation Group improved perturbative calculation.
We will study a two dimensional quantum field theory off criticality. In the 
two scaling limits, the infrared (IR) and the ultraviolet (UV), it becomes a
conformal field theory characterised by the Virasoro central charge.  
The field theory we look at has the 
minimal models ${\cal M}(m)$ as scaling limits.
Our approximation method obtains a value for the
infrared central charge
$c_{IR}$ using perturbative information around the ultraviolet limit of the
theory. 
In \cite{zam2,cardy2,laessig} $c_{IR}$ was calculated in perturbation
theory in the limit when $m\rightarrow \infty$. This is valid because
then the UV and IR fixed-points are arbitrarily close in the
coupling constant space. The renormalisation group (RG)
eigenvalue $y=\tfrac{4}{m+1}$ of the perturbation vanishes as 
$m\rightarrow \infty$ 
and the perturbation thus becomes marginal, and hence the field theory does
not move away from the UV fixed-point.
As well as for large values of $m$ our approximation also applies for small 
values, $(m>10)$, and we
can test it against the exact result $c_{IR}=c_{{\cal M}(m-1)}$ when
$c_{UV}=c_{{\cal M}(m)}$.
In the limit $m \rightarrow \infty$ we obtain the perturbative result
reported in \cite{zam2,cardy2,laessig}. 
To improve upon standard perturbative techniques requires some additional
input. For this we will exploit the analyticity of the
energy-momentum correlator $\langle TT \rangle$ in a complex scale
parameter using it to construct a dispersion relation.
In section 2 we describe this dispersion relation. Section 3 gives the
results from perturbative CFT and the RG improvements, and in section 4 
we illustrate our approximation with numerical results.

\section{The dispersion relation}
We will now construct a dispersion relation that relates the infra-red 
and ultra-violet behaviour of the energy-momentum tensor
two-point function.
The correlator of two energy-momentum tensors can be written using
the K\"{a}llen-Lehman  spectral representation \cite{cappelli} as
\begin{equation}
  \label{spectral}
\langle T_{zz}(z,\bar{z})T_{zz}(0,0)\rangle = 
\frac{\pi}{48}\int_0^\infty  d\mu \, \tilde{c}(\mu)
\int \frac{d^2 p}{(2\pi)^2} \frac{e^{\frac{i}{2}(p\bar{z}+\bar{p}z)}}
{p\bar{p}+\mu^2} \, \bar{p}^4\equiv \frac{F(|z|^2)}{2\, z^4}
\nonumber
\end{equation}
where we use the usual complex variables $z,\bar{z}$, and
$\tilde{c}(\mu)d\mu$ is the spectral density of the QFT which represents the 
density in degrees of freedom at the mass $\mu$. 
Integrating over $p$ gives
\begin{equation}
  \label{F2}
 F(s) =  \frac{1}{48}\int_0^\infty\! \! \! d\mu \, \tilde{c}(\mu)
 \mu \sqrt {s}\left( ({\mu}^{3}{s}^{3/2}\!+\!24\mu 
\sqrt{s}) K_0(\mu \sqrt{s})\!+\!(8 \mu^2 s\!+\!48) K_1(\mu \sqrt {s})\right )
\end{equation}
where $s\in {\mathbb R}_+$ and $K_0,K_1$ are the modified Bessel functions.
In \cite{cappelli} it was shown using the spectral representation that
$c_{UV}=\int_0^\infty d\mu \tilde{c}(\mu)$ and $c_{IR}=\lim_{\epsilon
  \rightarrow 0}\int_0^\epsilon d\mu \tilde{c}(\mu)$. 
 From the properties of the modified Bessel functions it then follows from
 \eqref{F2} that $F(s)$ has the limits 
\begin{equation}
\label{Fgraense}
F(s)\rightarrow \left\{ 
\begin{array}{ll}
c_{UV} & \mbox{for }s\rightarrow 0_{+}, \\
c_{IR} & \mbox{for }s\rightarrow \infty. 
\end{array} \right.
\end{equation}
In \cite{key} we prove in detail that
the expression \eqref{F2} provides an analytic continuation 
from real positive values
of $s$ to the complex plane cut along the negative real axis, 
as might be expected from the analyticity of the Bessel functions themselves.
From now on we take $s$ to be in the complex plane with a thin wedge about
the negative real axis removed (so $-\pi+\epsilon < \arg(s)<\pi-\epsilon$).
From \eqref{F2} it also follows that $F(s)$ has the limits \eqref{Fgraense}
for all $s\rightarrow 0$ and $|s| \rightarrow \infty$ from the cut complex
plane, so $\lim_{|s|\rightarrow \infty}F(s)=c_{IR}$ and $F(0)=c_{UV}$.

The following contour integral in the cut complex plane therefore vanishes by
analyticity of $F(s)$
\begin{equation}
  \label{contourint2}
  \frac{1}{2\pi i}\int_C ds \, \frac{e^{\rho/ s}}{s}F(s^{})=
\frac{1}{2\pi i}\left(\int_{C_0}ds+\int_{C_1}ds+\int_{C_2}ds\right)
\frac{e^{\rho/ s}}{s}F(s^{})=0,
\end{equation} 
where the contour $C$ is given in figure 1. 
\begin{figure}
\epsfxsize=8cm
\epsfysize=8cm
\begin{center}
\epsffile{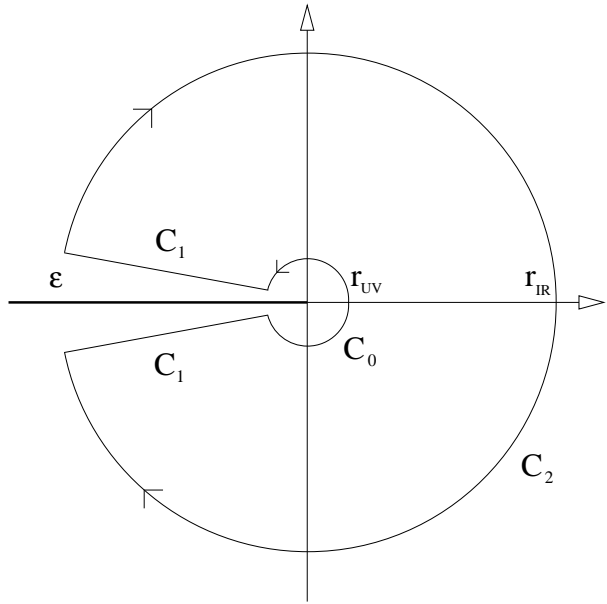}
\end{center}
\caption{Integration contour C in the cut complex plane.}
\label{fig:contour}
\end{figure} 
The contribution from the large circle 
$C_2$, where $s=r_{IR}e^{i\theta}$ for $\theta \in [\pi-\epsilon,-\pi+
\epsilon ]$, picks out the value
$(-1+\tfrac{\epsilon}{\pi})\lim_{|s|\rightarrow \infty}F(s)=(-1+\tfrac{\epsilon}{\pi})c_{IR}$ in the limit $r_{IR}
\rightarrow \infty$. This is seen by writing the integral as an angular
integral over $\theta$, the limit $r_{IR}\rightarrow \infty$ can then be
taken before the integral as it follows from \eqref{F2} that the integrand is
bounded by a constant in the cut plane.
For $\epsilon$ small the infrared central charge can thus be calculated from an
integral over the small circle $C_0$ and an integral over the contour $C_1$
which we denote Cut$(\rho)$
\begin{equation}
  \label{cir10}
  c_{IR}=\frac{1}{2\pi i}\int_{C_0}ds\ \frac{e^{\rho/
  s}}{s}F(s^{})+\mbox{Cut}(\rho)=I(\rho)+ \mbox{Cut}(\rho).
\end{equation}
The exponential factor in the integrand suppresses the contribution of the 
cut for large positive values of $\rho$ and 
$\lim_{\rho \rightarrow \infty}\mbox{Cut}(\rho)=0$ so that 
$c_{IR}=\lim_{\rho\rightarrow\infty}I(\rho)$. To show this we divide the
$|s|$ interval into $[r_{UV},1]$ and $[1,r_{IR}]$ and note that $F(s)$ is finite in
the unit disk hence the contribution in the lower region is bounded by the
function $k\tfrac{e^{-\rho}}{\rho}$ for some $k\in \mathbb{R}_+$. In the
other region, using the properties of the Bessel functions, the integral is
seen to be bounded by the expression $k \int d\mu
\tilde{c}(\mu)\mu^{7/2}\int_1^{r_{IR}}dr e^{-\rho \cos(\e)/r}r^{3/4}e^{-\mu
    \sqrt{r}\sin{\e/2}}$ which is uniformly convergent in $\rho$, the
details are shown in \cite{key}.  

For small $s$, $F(s)$ and hence $I(\rho)$ are determined by the ultra-violet
behaviour. Close to the UV fixed-point we may describe
this by perturbation theory with running coupling constant $\bar{g}(s)$,
vanishing as $s\rightarrow 0$. We denote the
perturbative approximation to the integral in \eqref{cir10} by $I_n(\rho)$ 
where
$F(s^{})$ is approximated to n-th order by the perturbative expression
$F_n(s^{})$. If analyticity is not spoilt by the perturbative expansion (as
may be seen by inspection of the results in the next section) then simply 
substituting $I_n$ for $I$ in this limit would yield 
$c_{IR}\approx\lim_{\rho\rightarrow\infty}I_n(\rho)=
\lim_{s\rightarrow\infty}F_n(s^{})$,
which is just the perturbative estimate of the central charge, $c_{IR}^*$.
Since the scale dependence of the theory can be absorbed in the
scaling of the running coupling constant $\bar{g}(s)$ we can write 
$F_n(s)=\Phi_n(\bar{g}(s))$, so that if $g_{IR}^*$ is the first non trivial
zero of the perturbative $\beta$-function i.e.\ 
$\bar{g}(s)\rightarrow g_{IR}^{*}$ for $s\rightarrow \infty$,
then $\lim_{\rho\rightarrow\infty}I_n(\rho)=\Phi_n(g_{IR}^*)$.
However, we wish to improve on this result. 
Both $I_n(0)$ and $I(0)$ equal $c_{UV}$.
For small enough $\rho$,
$I_n(\rho)$ provides a good approximation to $I(\rho)$,
since the power series expansion of $I(\rho)$ is controlled by
the small $s$ expansion of $F(s)$ for which perturbation theory is good.
For larger values of $\rho$, $I(\rho)$ and $I_n(\rho)$  part company,
tending to $c_{IR}$ and $c_{IR}^*$ respectively. If $c_{IR}<c^*_{IR}$
and if the region where $I_n(\rho)$ 
is a good approximation to $I(\rho)$ is large enough, then
$I_n(0)-I_n(\rho)$ will have a maximum before approaching
its limiting value of $c_{UV}-c^*_{IR}$. It is this maximum
that we will use to provide a better estimate of $c_{UV}-c_{IR}$,
since this occurs at the largest value of $\rho$ for which $I_n(\rho)$
is a reasonable approximation to $I(\rho)$ and the true value of $c_{UV}-c_{IR}$
is given by $I(0)-I(\infty)$.
As we will see this is the case for the minimal models considered in
the next section.


\section{The perturbative calculation}
In the CFT ${\cal M}(m)$ the primary field $\phi_{(1,3)}$ is a relevant field
which RG trajectory is a geodesic in the coupling constant space
\cite{laessig}. The QFT
\begin{equation}
S=S_{{\cal M}(m)}-\lambda_0 \int d^2x\ \phi_{(1,3)}(x).
\label{13virkning}
\end{equation}
therefore interpolates between the UV fixed-point ($\lambda_0=0$) given by 
${\cal  M}(m)$ and the IR fixed-point which is given by the CFT 
${\cal M}(m-1)$. This
can be seen from perturbative arguments in the $m\rightarrow \infty$ limit
\cite{zam2,cardy2,cappellipert} and the general case is argued for in 
\cite{alzam}
using thermo-dynamical Bethe ansatz  methods.

When $\lambda_0$ is small we can use perturbation theory to calculate
quantities in the theory \eqref{13virkning}. Calling the bare field 
$\phi_{(1,3)}=\phi$ one
gets \cite{cappellipert} to order $\lambda_0$
\begin{eqnarray}
\langle \phi(x) \phi(0) \rangle &=&
\frac{\langle\phi(x)\phi(0)e^{\lambda_0\int d^2x' \phi(x')} \rangle_{{\cal M}(m)}}{\langle
  e^{\lambda_0\int d^2x'  \phi(x')}\rangle_{{\cal M}(m)}}\nonumber \\
&=&\frac{1}{| x |^{2(2-y)}}\left(1+\lambda_0\frac{4\pi b(y)
     A(y)}{y}| x |^y+O(\lambda_0^2)\right),
\nonumber
\end{eqnarray}
where  
$A(y)=\frac{\Gamma (1-y)\Gamma (1+y/2)^2}{\Gamma (1-y/2)^2 \Gamma (1+y)} 
=1+O(y^3)$ and $b(y)$ is the operator product expansion coefficient $C_{\phi \phi}^\phi$ which
can be calculated in the Coulomb gas representation of
minimal models using the formulas in \cite{dotsenko}. $b(y)$
is given by
\begin{equation}
\label{bi2}
b(y)^2=\tfrac{16}{3}\tfrac{(1-y)^4}{(1-y/2)^2(1-3y/4)^2}
\left(\tfrac{\Gamma(1+y/2)}{\Gamma(1-y/2)}\right)^4\left(
\tfrac{\Gamma(1-y/4)}{\Gamma(1+y/4)}\right)^3\left(\tfrac{\Gamma(1-y)}{\Gamma(1+y)}\right)^2\left(\tfrac{\Gamma(1+3y/4)}{\Gamma(1-3y/4)}\right)=
\tfrac{16}{3}+O(y).
\nonumber
\end{equation}
$y=\tfrac{4}{m+1}$ is the RG eigenvalue
for the perturbation $\phi_{(1,3)}$.
With the renormalisation conditions $\langle
\phi(x,g)\phi(0,g)\rangle|_{\mid x \mid=\mu^{-1}}\equiv \mu^4$, the
renormalised correlator and $\beta$-function, $\beta(g)\equiv \mu
\tfrac{dg(\mu)}{d\mu}$, becomes to the same order \cite{cappellipert}
\begin{eqnarray}
\langle \phi(x,g)\phi(0,g)\rangle &=&\frac{\mu^4}{|\mu x
  |^{2(2-y)}}\left(1+\frac{4\pi A(y) b(y) g}{y}(| \mu x
  |^y-1)\right),\nonumber \\  
 \beta(g) &=& -yg-\pi b(y) g^2 A(y)+O(g^3).
\label{phiphi}
\nonumber
\end{eqnarray}
Here $g$ and $\phi(x,g)$ are the renormalised coupling and field.
The RG fixed-points, the zeros of the $\beta-$function, are thus $g_{UV}=0$, $g_{IR}^*=\tfrac{-y}{\pi A(y)b(y)}$ and
therefore $g\in (\tfrac{-y}{\pi A(y)b(y)},0 )$ as the theory
\eqref{13virkning} lies between the two scaling limits.
The trace of the energy-momentum operator $\Theta$ is the infinitesimal
generator for scale transformations and it is given by \cite{zam2}
$\Theta(x)=2\pi \beta \phi(x)$.
Then it follows from  the Ward identities that
\begin{equation}
\begin{split}
\p_{\bar{z}_1}\p_{\bar{z}_2}\langle
T(z_1,\bar{z}_1)T(z_2,\bar{z}_2)\rangle
=\, & \frac{1}{4^2}\partial_{z_1}\partial_{z_2}\langle
\Theta(z_1,\bar{z}_1)\Theta(z_2,\bar{z}_2)\rangle \\ 
=\, & 
\frac{\pi^2}{4}\beta(g)^2\p_{z_1}\p_{z_2}\langle \phi (z_1,\bar{z}_1) 
\phi (z_2,\bar{z}_2) \rangle.
\label{diffeq1}
\end{split}
\end{equation}
Writing $\tilde{F}(\tilde{R})=2 z^4 \langle
T(z,\bar{z})T(0,0)\rangle$ in terms of the dimensionless variable
$\tilde{R}=\mu^2 z \bar{z}$, then \eqref{diffeq1} becomes
\begin{equation}
  \label{diffeq5}
  \frac{\p ^2}{\p \tilde{R}^2}\tilde{F}(\tilde{R})=\frac{\pi^2 \beta^2}{4\mu^4}
  \tilde{R}^2 \frac{\p^2}{\p \tilde{R}^2}\langle \phi(\tilde{R}) \phi(0) 
\rangle .
\end{equation}
The limiting values of $\tilde{F}(\tilde{R})$ are given by \eqref{Fgraense},
using these the solution to \eqref{diffeq5} becomes
\begin{equation}\label{intphiphi}
\begin{split}
 \tilde{F}(\tilde{R})=&\
c_{UV}+\frac{\pi^2 g^2
  \tilde{R}^y}{2}\left(\frac{y(2-y)(3-y)}{y-1}+2\pi
  A(y)b(y)g\left( \frac{(2-y)(3-y)}{1-y} \right. \right. \\ 
& \left. \left. +\tilde{R}^{\frac{y}{2}}\frac{(3y-4)(3y-6)}{3(\frac{3}{2}y-1)}\right) \right).
\nonumber
\end{split}
\end{equation}
The function $F(s)$ is determined as $F(s)=
\tilde{F}(\tilde{R})|_{\tilde{R}=1,g=\bar{g}(s)}$. The theory is thus fixed
at the renormalisation point and all scale dependence is moved into the running
coupling constant \cite{cappellipert}
\begin{equation}
  \label{runnings}
  \bar{g}(s)=\frac{g s^{\frac{y}{2}}}{1-\frac{\pi
  A(y)b(y)g}{y}(s^{\frac{y}{2}}-1)}. 
\end{equation}
We thereby get our 1 loop RG improved approximation of $F(s)$
\begin{equation}  \label{approxf}
\begin{split} 
 F_1(s)=&\ c_{UV}+\frac{\pi^2}{2}\bar{g}^2(s)\left(
\frac{y(2-y)(3-y)}{y-1} \right. \\ 
&\left.\ +2\pi A(y)b(y)\bar{g}(s)\left(
\frac{(2-y)(3-y)}{1-y}+\frac{(3y-4)(3y-6)}{3(\frac{3}{2}y-1)} \right) \right).
\end{split}
\end{equation}
To apply our approximation based on the dispersion relation of the previous 
section to calculate $\Delta c=c_{UV}-c_{IR}$ we now have to find the maximum
of 
\begin{equation}
I_1(0)-I_1(\rho)=c_{UV}-\frac{1}{2\pi i}\int_{C_0}ds
\frac{e^{\rho/ s}}{s}F_1(s^{}).
\end{equation}

\section{Results}
We first note that $F_1(s)$ has the correct value in the perturbative limit
$y\rightarrow 0$. From \eqref{approxf} it follows that $\Delta c_{pert}=
c_{UV}-c_{IR}^*=\lim_{s\rightarrow 0}(c_{UV}-F_1(s^{}))$ is given by
\begin{equation}\label{delcpert}
\begin{split}
\Delta c_{pert} 
 =&\, - \frac{\pi^2}{2}(g_{IR}^*)^2\left(
\frac{y(2-y)(3-y)}{y-1} \right. \\ 
&\left.\ +2\pi A(y)b(y)g_{IR}^*\left(
\frac{(2-y)(3-y)}{1-y}+\frac{(3y-4)(3y-6)}{3(\frac{3}{2}y-1)} \right) \right)
 \\ 
=&\, \frac{3y^3}{16}+O(y^4),
\end{split}
\end{equation}
and the exact value is 
\begin{equation}
\label{exactval}
\Delta c_{exact}=c(m)-c(m-1)=
\frac{12}{m(m^2-1)}=\frac{3y^3}{2(2-y)(4-y)}=\frac{3y^3}{16}+O(y^4).
\end{equation}
To calculate $\Delta c$ for finite $m$ we have to
calculate $I_1(\rho)$. The running coupling \eqref{runnings} can be rewritten
as 
\begin{equation}
  \label{runnings2}
  \bar{g}(s)=\frac{g s^{\frac{y}{2}}}{1-\frac{\pi
  A(y)b(y)g}{y}(s^{\frac{y}{2}}-1)}=\frac{g_{IR}^*
  |\tilde{g}|s^{y/2}}{1+|\tilde{g}|s^{y/2}},\ \ \ \ \ \ 
\tilde{g}=\frac{g}{g-g_{IR}^*}\ \in \ \ ( -\infty,0). 
\end{equation}
A variable change $s'=s{|\tilde{g}|^{2/y}}$ in $I_1(\rho)$ then leads
to 
\begin{equation}
  \label{runningc2}
  I_1(\rho,g)=\frac{1}{2\pi i}\int_{C_0}ds \frac{e^{\rho/ s}}{s}F_1(s^{})=
\frac{1}{2\pi i}\int_{C'_0}ds' 
\frac{e^{\rho'/s'}}{s'}F_1({s'})=\tilde{I}_1(\rho')
\end{equation}
where now all dependence of the renormalised coupling $g$ is moved into
$\rho'=\rho |\tilde{g}|^{2/y}$ and the radius of $C'_0$:
$r'_{UV}=r_{UV}{|\tilde{g}|^{2/y}}$ as now
$\bar{g}(s')=\tfrac{g_{IR^*}}{1+{s'}^{y/2}}$. 
\begin{figure}
\epsfxsize=14cm
\epsfysize=8cm
\begin{center}
\epsffile{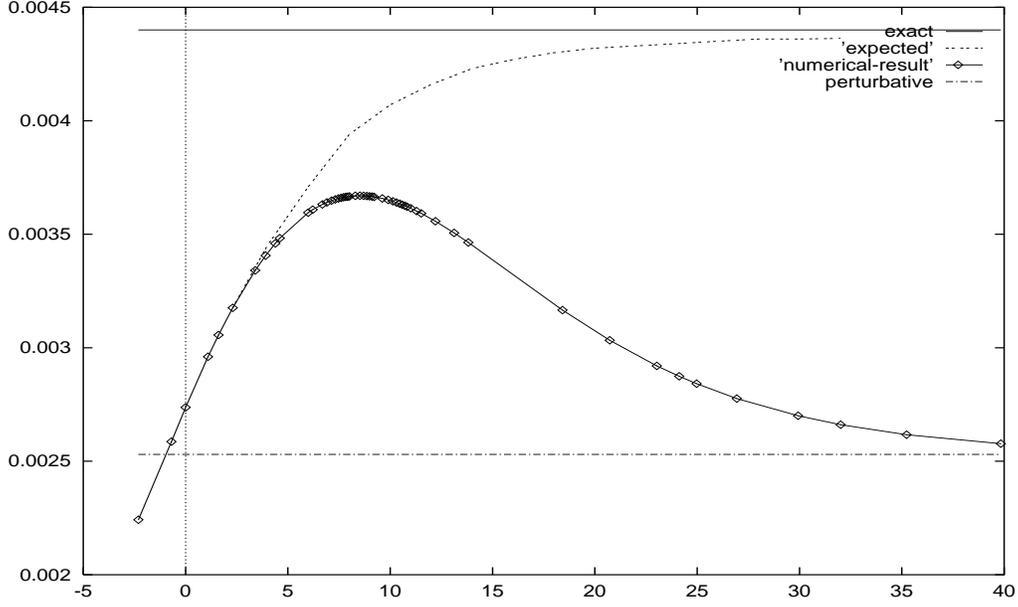}
\end{center}
\caption{The numerical-result $\tilde I_1(0)-\tilde I_1(\rho')$ against
  $\log{\rho'}$, also plotted is the exact value $\Delta c_{exact}$ 
and the RG improved perturbative result $\Delta c_{pert}$ all for $m=14$. 
The dashed line is the expected behaviour of $\tilde{I}(0)-\tilde{I}(\rho')$.}
\label{fig:resulm14}
\end{figure} 
The integral $\tilde I_1(0)-\tilde I_1(\rho')$ can be calculated numerically
and in figure 2 we have plotted it against $\log \rho'$ for the case
of $m=14$. For large $\rho'$ it tends to the perturbative value $0.0025$,
but has a maximum value of $0.00367$ which provides a better approximation to
the true value of $0.00440$. The dashed line indicates how we expect 
$\tilde{I}(0)-\tilde{I}(\rho')$ to behave. We have approximated $\Delta c$
for different values of $m$ by the maximum
of $\tilde I_1(0)-\tilde I_1(\rho')$ and we denote this approximation as 
$\Delta c_{approx}$.
The numbers $\Delta c_{approx}$ have been obtained by a numerical integration
using a NAG mark 18 Fortran Library quadrature routine.
In figure 3 the error in $\Delta c_{approx}$ and $\Delta c_{pert}$ 
compared with $\Delta c_{exact}$ is plotted against $m$. 
The errors are scaled with $m(m^2-1)$ so that all points are
distinguishable on the same plot. The figure shows that the error in the
approximation $\Delta c_{approx}$ is more than a factor 2 smaller than
the error in the perturbative value $\Delta c_{pert}$ in the region plotted. 

For values of $m$ smaller than $11$ perturbation theory begins to breakdown 
indicated by the RG improved result turning negative (violating unitarity).
Our approximation whilst still positive for $m$ close to $10$ also becomes 
poorer since it is based on the RG expression. 
Also, note that both $\Delta c_{approx}$ and $\Delta c_{pert}$ 
approaches $\Delta c_{exact}$ faster than the asymptotic value 
$3y^3/16$ in the limit $m\rightarrow \infty$.
\begin{figure}
\epsfxsize=13cm
\epsfysize=7cm
\begin{center}
\epsffile{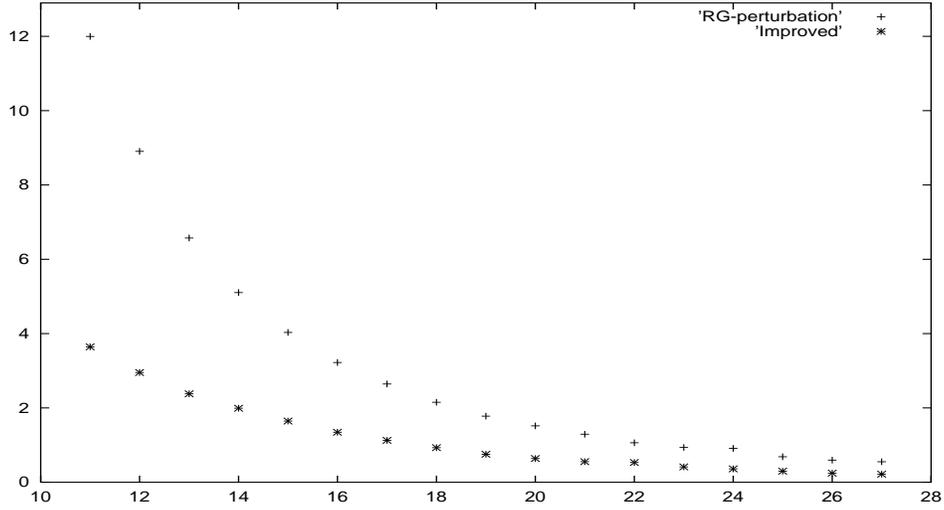}
\end{center}
\caption{$(\Delta c_{exact}-\Delta c_{approx})m(m^2-1)$ and $(\Delta c_{exact}-\Delta c_{pert})m(m^2-1)$ 
against  $m$.}
\label{fig:error}
\end{figure}

\section{Conclusion}
We have described an approximation method to obtain the IR central charge of
the minimal models valid for $m>10$. The method improves upon the
RG improved calculation by exploiting the analyticity
of the
correlator of the energy-momentum tensor. This
analyticity is a property both of the exact correlator and
the RG improved perturbative estimate of it, \cite{key}, which we have used 
in our calculation.

Our approximation is correct in the perturbative limit $m \rightarrow \infty$. 
For smaller $m$ we obtained the values shown in figure 3 
together with the RG improved perturbative
result. This figure demonstrates that the approximation is significantly 
better than the RG 
improved perturbative result, e.g.\ for  $m=18$ it has a
7.8\% relative deviation from the exact result whereas the RG perturbative
result deviates by 18.0\%. 
The analyticity in the complex scale parameter $s$ of the energy-momentum 
tensor two-point function is in fact an ubiquitous property of correlation
functions in quantum field theory, having its origin 
in the hermiticity of the Hamiltonian.
Consequently we expect our approach to be applicable beyond the specific
calculation we have used to illustrate it here, to other RG improved 
perturbative calculations of Green's functions.\\
\ \\
{\large {\bf Acknowledgement}}\\
Lars Kj\ae rgaard acknowledges a research grant from the Danish Research
Academy.

\end{document}